\documentclass[prb,twocolumn,showpacs,preprintnumbers,
amsmath,amssymb,superscriptaddress]{revtex4}

\usepackage{graphicx}
\usepackage{bm}

\newcommand{\beq}{\begin{equation}}
\newcommand{\eeq}{\end{equation}}
\newcommand{\beqa}{\begin{eqnarray}}
\newcommand{\eeqa}{\end{eqnarray}}

\newcommand{\w}{\omega}

\newcommand{\ket}[1]{\left| #1 \right\rangle}

\begin{document}

\title{Anticrossings in F\"orster coupled quantum dots}
\author{Ahsan~Nazir}
\email{ahsan.nazir@materials.oxford.ac.uk}
\affiliation{Department of Materials, Oxford University, Oxford OX1 3PH, United Kingdom}
\author{Brendon~W.~Lovett}
\affiliation{Department of Materials, Oxford University, Oxford OX1 3PH, United Kingdom}
\author{Sean~D.~Barrett}
\affiliation{Hewlett-Packard Laboratories, Filton Road, Stoke Gifford, Bristol BS34 8QZ, United Kingdom}
\author{John~H.~Reina}
\altaffiliation[On leave of absence from ]{Centro Internacional de F\'{\i}sica (CIF), A.A. 4948, Bogot\'a, Colombia}
\affiliation{Department of Materials, Oxford University, Oxford OX1 3PH, United Kingdom}
\affiliation{Centre for Quantum Computation, Clarendon Laboratory, Department of Physics, Oxford University, Oxford OX1 3PU, United Kingdom}
\author{G.~Andrew~D.~Briggs}
\affiliation{Department of Materials, Oxford University, Oxford OX1 3PH, United Kingdom}
\date{\today}

\begin{abstract}
We consider two coupled generic quantum dots, each
modelled by a simple potential which
allows the derivation of an analytical expression for the inter-dot
F\"orster coupling, in the dipole-dipole approximation. We investigate the energy level behaviour of this coupled two-dot system under the influence of an external applied electric field and predict the presence of anticrossings in the optical spectra due to the F\"orster interaction.
\end{abstract}


\maketitle

\section{\label{intro}Introduction}

Excitons (electron-hole bound states) within quantum dots (QD's) have
already attracted much interest in the field of quantum
computation and have formed the basis of several proposals for quantum logic
gates.~\cite{qdqc02} The energy shift due to the exciton-exciton dipole
interaction between two QD's gives rise to diagonal terms in the interaction
Hamiltonian, and hence it has been proposed that quantum logic may be
performed via ultra-fast laser pulses.~\cite{biolatti02}
However, excitons within adjacent QD's are also able to
interact through their resonant (F\"orster) energy
transfer,~\cite{forster59} some evidence for which has been obtained experimentally
in a range of systems.~\cite{kagan96a,crooker02,berglund02,hu02} As is shown below, this resonant transfer of energy gives
rise to off-diagonal Hamiltonian matrix elements and therefore to a naturally
entangling quantum evolution. It is proposed here that this interaction may be observed in a straightforward way through the observation of anticrossings induced in the coupled dot energy spectra by the application of an external static electric field. We also show that the off-diagonal matrix elements can be made sufficiently large to be of interest for excitonic quantum computation.~\cite{quiroga99,lovett02a}

The first studies of F\"orster resonant energy transfer were performed in
the context of the sensitized luminescence of
solids.~\cite{forster59,dexter53} Here, an excited sensitizer atom
can transfer its excitation to a neighbouring acceptor atom, via
an intermediate virtual photon. This same mechanism has also been
shown to be responsible for exciton transfer between QD's,~\cite{crooker02} and within molecular systems~\cite{berglund02}
and biosystems~\cite{hu02} (though incoherently, as a mechanism for photosynthesis), all of which may be treated in a similar formulation.
Thus, the results reported here can be expected to apply not only to QD systems, but also to a wide range of nanostructures where F\"orster processes are of primary significance.

In this paper, we consider two coupled generic QD's, each
modelled by a simple potential which is given by infinite
parabolic wells in all three ($x$, $y$, $z$) dimensions. This
potential profile allows an analytical expression for the interdot
F\"orster coupling in the dipole-dipole approximation to be derived and,
although we expect it only to allow for qualitative predictions about real observations, similar models have been successfully used in the literature.~\cite{krummheuer02,damico01,jacak98} Excitations of each dot are assumed
to be produced optically, and we neglect tunneling effects between
the two coupled dots (see the Appendix for more details). We shall restrict ourselves to small, strongly confined dots, with electron and hole confinement energies $\sim100$~meV and low temperatures ($T<5$~K; see Sec.~\ref{decay}).
We shall therefore consider only the ground state (no
exciton) and first excited state (one exciton) within our model, as they will be energetically well separated from higher excitations. These two states define our qubit basis as $\ket{0}$ and $\ket{1}$, respectively.

Various techniques exist for growing QD's in the
laboratory, of which the Stranski-Krastanow method~\cite{eaglesham90,xie95} is possibly the most promising for
the realization of a controllably coupled many-dot system. In this growth
mode a semiconductor is grown on a substrate which is made of a different semiconductor,
leading to a lattice mismatch between the layers. Under certain growth conditions, dots form
spontaneously due to the competing energy considerations of dot
surface area, strain, and volume. If a spacer layer
of material is then grown above the first dot layer and then a second dot layer deposited, a vertically
correlated arrangement of dots can be made.~\cite{tersoff96} Two
such stacked dots could then form our interacting two-qubit system
with materials, growth conditions, and spacer layer size tailored
to give suitable electronic properties and inter-dot Coulomb
interactions.

\section{\label{tqdm}Quantum Dot Model}
\subsection{Single-particle states}
Many varying approaches to the calculation of electron and
hole states in QD's have been put forward in Refs.~\onlinecite{biolatti02,krummheuer02,damico01,jacak98,califano99,gang97a,krishna91,wang94,franceschetti97}, the choice of which depends on the final aim of the work. The aim here is to provide a clear and simple illustration of how to experimentally observe resonant energy transfer between a pair of QD's and how to exploit
this inter-dot interaction to perform quantum logic. Therefore, we shall consider one of the most basic models, similar to those in Refs.~\onlinecite{krummheuer02,damico01,jacak98}, which treats the conduction- and valence-band ground states as those of a three-dimensional infinite parabolic well. This simple model can provide us with
analytical expressions for the energies and wavefunctions of the single-particle
states and for the dipole-dipole interaction between two dots. We do not expect that this model will predict the precise values of experimental measurements; however, it is expected to indicate qualitative trends in real observations.

In the effective mass and envelope function
approximations~\cite{bastard81,harrison00} the Schr\"{o}dinger
equation for single particles may be written as
\begin{equation}\label{schrodinger}
   H_i({\bf r})\phi_i({\bf r})= \left[-\frac{\hbar^2}{2}\nabla\left(\frac{1}{m_i^*}\right)\nabla+V_i({\bf r})\right]\phi_i({\bf r})
    = E_i\phi_i({\bf r}),
\end{equation}
where $i=e, h$ for electron or hole, $V_i({\bf r})$ is the dot
confinement potential which accounts for the difference in band gaps
across the heterostructure, and $m_i^*$ is the effective mass
of particle $i$. Here, $\phi_i(\bf{r})$ is the envelope function
part of the total wavefunction:\begin{equation}\label{envelope}
\psi_i({\bf r}) = \phi_i({\bf r}) U_i({\bf r}).
\end{equation}
The envelope function describes the slowly varying contribution to
the change in wavefunction amplitude over the dot region, and the
physical properties of the single-particle states can be derived
purely from this contribution. $U_i({\bf r})$ is called the Bloch
function and has the periodicity of the atomic lattice. Its
consideration is vital when describing the interactions between
two or more particles.

\begin{figure}[t]
\centering
\includegraphics[width=3.8in,height=2.5in]{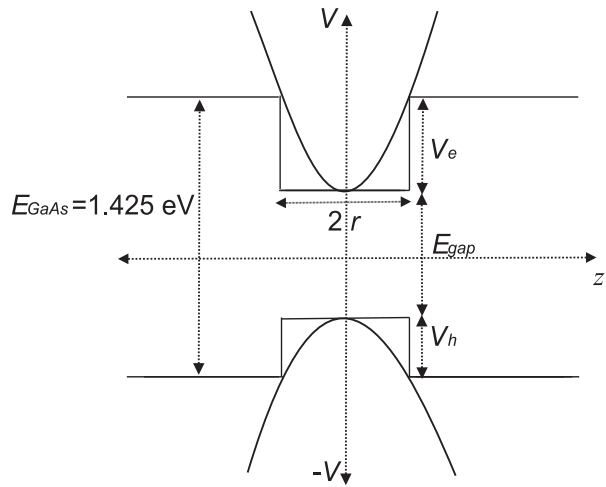}
\caption[Choice of dot parameters]{The parameters $c_j$ are chosen from electron and hole confinement potentials $V_e$ and $V_h$ by taking $(1/2)c_jr^2=V_i$, for $j=x,y,z$, and $i=e,h$. This matches a well of depth $V_i$ with the parabolic potential $(1/2)c_jj^2$ at a width $r$ from the dot centre. All parameters are chosen to be consistent with GaAs spacer layers, i.e. $V_e+V_h+E_{gap}=1.425$~eV, the GaAs band-gap.}
\label{dotparameters}
\end{figure}

In the approximate analytical model, a separable potential comprising
infinite parabolic wells in all three dimensions represents the
QD:\footnote{With this choice of potential we are able to model both the convenient situation of a spherically symmetric QD, and the more common situation in self-assembled dots of stronger confinement in the growth ($z$) direction than in the $x, y$ plane.}
\begin{equation}\label{parabolicpotn}
    V(x, y, z) = \frac{1}{2}c_{i,x}x^2+\frac{1}{2}c_{i,y}y^2+\frac{1}{2}c_{i,z}z^2,
\end{equation}
where the frequency $\omega_{i,j}=\sqrt{c_{i,j}/m^*}$ for $j=x, y ,z$ (see Fig.~\ref{dotparameters}).
Hence, the Schr\"{o}dinger
equation~(Eq.~\ref{schrodinger}) is also separable and provides simple
product solutions for the electron and hole states.
The envelope functions are therefore given by
\begin{equation}\label{separationparabolic}
    \phi_i({\bf r}) = \xi_{i,x}(x)\xi_{i,y}(y)\xi_{i,z}(z),
\end{equation} for the parabolic confinement $({\bf r} = (x, y, z))$.
We now drop the subscript $i$ but remember that due to their differing
effective masses electrons and holes may take different values for
the parameters defined throughout this paper.

The solutions to the one-dimensional Schr\"{o}dinger equation for
the potential form of Eq.~\ref{parabolicpotn} are given
by:
\begin{equation}\label{parabolicsols}
\xi^n(x)=\left(\frac{1}{n!2^nd_x\sqrt{\pi}}\right)^{1/2}H_n\left(\frac{x}{d_x}\right)\exp\left({-\frac{x^2}{2d_x^2}}\right),
\end{equation}
in the $x$- direction with analogous expressions for $y$ and $z$.
The integer $n=(0,1,2,3,...)$ labels the quantum state, with
energy $E_n=(n+1/2)\hbar\omega_x$, the $H_n$'s are Hermite
polynomials, and
$d_x=\left(\hbar/\sqrt{m^*c_x}\right)^{1/2}=\left(\hbar/(m^*\omega_x)\right)^{1/2}$.
We are interested only in the ground state solutions of each well,
so our envelope function is given by:
{\setlength\arraycolsep{0.0em}
\begin{eqnarray}\label{envelopeparabolic}
\phi(x, y, z)&{}={}&\left(\frac{1}{d_xd_yd_z\pi^{3/2}}\right)^{1/2}\exp\left({-\frac{x^2}{2d_x^2}}\right)
\nonumber\\
  & & {\times}\:\exp\left({-\frac{y^2}{2d_y^2}}\right)\exp\left({-\frac{z^2}{2d_z^2}}\right),
\end{eqnarray}}with energy $E_0=\frac{1}{2}\hbar(\omega_x+\omega_y+\omega_z)$.
The choice of constants $c_j$ and hence $d_j$ will be different for changing confinement potentials
and particle masses, and so will depend upon the
energies of the system under consideration and whether the
particle is an electron or hole (see Fig~{\ref{dotparameters}}).

\subsection{Excitons and Coulomb integrals}

The excitation of an electron from a valence band state to a
conduction band state leaves a hole in the valence band. The
electron and hole are oppositely charged and may form a bound
state, the exciton, with the absence or presence of a
ground state exciton within a dot forming our qubit basis
($\ket{0}$ and $\ket{1}$ respectively). For excitons, we must
consider an electron-hole pair Hamiltonian
\begin{equation}\label{ehpairH}
    H=H_e+H_h-\frac{e^2}{4\pi\epsilon(\bf r_e-\bf r_h)|\bf r_e-\bf r_h|}+E^{gap},
\end{equation}
where $H_e$ and $H_h$ are given by Eq.~\ref{schrodinger} with the
appropriate effective masses and potentials; $E^{gap}$ is the
semiconductor band-gap energy, and $\epsilon(\bf r_e-\bf r_h)$ is
the background dielectric constant of the semiconductor. We shall
consider the simplest case of $\epsilon({\bf r_e}-{\bf
r_h})=\epsilon_0\epsilon_r$, i.e. the relative
permittivity $\epsilon_r$ is independent of $(\bf r_e-\bf r_h)$. The intra-dot
energy shift due to the Coulomb term $H_{eh}=e^2/4\pi\epsilon_0\epsilon_r|\bf r_e-\bf
r_h|$ is a small contribution to the total energy and we treat it
as a first-order perturbation. This strong confinement regime treats the electron and hole as independent particles with energy states primarily determined by their respective confinement potentials.~\cite{bryant88} It is valid for small dots with sizes less than the corresponding bulk exciton radius $a_0$ ($\sim 35$~nm~for InAs, $\sim13$~nm~for GaAs). For more sophisticated treatments of the calculation of excitonic states see, for example, Refs.~\onlinecite{biolatti02} (direct-diagonalization),~\onlinecite{franceschetti97} (psuedopotential calculations), and~\onlinecite{tranthoai90} (variational methods). In Ref.~\onlinecite{franceschetti97} it was found that a simple perturbation method was in good agreement with a self-consistent field approach.

We construct an antisymmetric wavefunction representing a single
exciton state given by
\begin{equation}\label{antisymmetricexciton}
    \Psi_{\rm I}= A\left[\psi_n'({\bf r_1,\sigma_1}),\psi_m({\bf r_2,\sigma_2})\right],
\end{equation}
where $\bf r$ and $\sigma$ are position (from the centre of the dot) and spin variables
respectively, $n$ and $m$ label the quantum states, and $A$ denotes overall antisymmetry. Here, one
electron $\psi_n'({\bf r_1,\sigma_1})$ has been promoted from the
valence band into a conduction band state whilst $\psi_m({\bf
r_2,\sigma_2})$ represents a state in the valence band.
Taking the Coulomb matrix element $\langle\Psi_{\rm F}|H_{eh}|\Psi_{\rm I}\rangle$ between
the initial state $\Psi_{\rm I}$ above and an identical state $\Psi_{\rm F}$ (in effect coupling an electon and hole via the Coulomb operator)
leads to two terms,~\cite{lovett02,franceschetti99} the direct
term
{\setlength\arraycolsep{0.0em}
\begin{eqnarray}\label{directeh}
M_{\rm IF}^{Direct}&{}={}&\frac{e^2}{4\pi\epsilon_0\epsilon_r}\int\int\psi_n'^*({\bf
r_1})\psi_n'({\bf r_1})\frac{1}{|\bf r_1-\bf r_2|}\nonumber\\
& & {\times}\:\psi_m^*({\bf r_2})\psi_m({\bf r_2})\textrm{d} \bf r_1 \textrm{d}
\bf r_2 ,
\end{eqnarray}}
and the exchange term
{\setlength\arraycolsep{0.0em}
\begin{eqnarray}\label{exchangeeh}
M_{\rm IF}^{Exch}&{}={}&\pm\frac{e^2}{4\pi\epsilon_0\epsilon_r}\int\int\psi_n'^*({\bf
r_1})\psi_m({\bf r_1})\frac{1}{|\bf r_1-\bf r_2|}\nonumber\\
&&{\times}\:\psi_n'({\bf r_2})\psi_m^*({\bf r_2})\textrm{d} \bf r_1 \textrm{d}
\bf r_2 .
\end{eqnarray}}
The sign of the exchange term is determined by the symmetry of the
spin state of the two particles; with the perturbation $H_{eh}$ being positive, triplet spin states give negative
exchange elements whereas singlet spin states give positive
values.
We shall now show how to calculate the direct electron-hole
Coulomb matrix element on a single dot where $n$ and $m$ are both
taken as ground states. The exchange interaction is much
smaller~\cite{franceschetti97,lovett02} and we shall not consider it here.

If we consider identical potentials in all three directions
then we can use the spherical symmetry to derive an
analytical expression for the direct Coulomb matrix element, which we call
$M_{eh}$. For $d_x=d_y=d_z=d$, Eq.~\ref{envelopeparabolic} may
be written in spherical polar coordinates as
\begin{equation}\label{parabolicpolar}
    \phi({\bf r}) = \left(\frac{1}{d\sqrt{\pi}}\right)^{3/2}\exp\left({-\frac{r^2}{2d^2}}\right).
\end{equation}
Substituting into Eq.~\ref{directeh} leads to
{\setlength\arraycolsep{0.0em}
\begin{eqnarray}\label{d1}
M_{eh}&{}={}&\frac{e^2}{4\pi\epsilon_0\epsilon_r}\left(\frac{1}{d_e\sqrt{\pi}}\right)^3
\left(\frac{1}{d_h\sqrt{\pi}}\right)^3\int\int
\exp\left({-\frac{r_1^2}{d_e^2}}\right)\nonumber\\
&&{\times}\:\exp\left({-\frac{r_2^2}{d_h^2}}\right)\frac{1}{|\bf
r_1-\bf r_2|}\textrm{d} \bf r_1 \textrm{d} \bf r_2,
\end{eqnarray}}where the contribution of the Bloch functions
$U(\bf r)$ has been neglected.~\cite{lovett02} We now express
$1/|\bf r_1-\bf r_2|$ in terms of Legendre polynomials
as~\cite{schiff55}
\begin{equation}\label{legendre}
     \frac{1}{|\bf r_1-\bf r_2|}=\left\{%
\begin{array}{ll}
\frac{1}{r_1}\sum_{l=0}^\infty(\frac{r_2}{r_1})^lP_l(\cos\theta), & \hbox{for $r_1> r_2$} \\
    \\
    \frac{1}{r_2}\sum_{l=0}^\infty(\frac{r_1}{r_2})^lP_l(\cos\theta), & \hbox{for $r_1< r_2$.} \\
\end{array}%
\right.
\end{equation}
Substituting this into Eq.~\ref{d1} and integrating over polar
angles leads to
{\setlength\arraycolsep{0.0em}
\begin{eqnarray}\label{d2}
M_{eh}&{}={}&\frac{4\pi e^2}{\epsilon_0\epsilon_r}\left(\frac{1}{d_e\sqrt{\pi}}\right)^3\left(\frac{1}{d_h\sqrt{\pi}}\right)^3\int_0^\infty \exp\left({-\frac{r_1^2}{d_e^2}}\right)r_1^2\textrm{d} r_1\nonumber\\
&&{\times}\:\bigg\{\int_0^{r_1}
\frac{1}{r_1}\exp\left({-\frac{r_2^2}{d_h^2}}\right)r_2^2\textrm{d}
r_2
\nonumber\\
&&{+}\:\int_{r_1}^\infty \frac{1}{r_2}\exp\left({-\frac{r_2^2}{d_h^2}}\right)r_2^2
\textrm{d} r_2\bigg\},
\end{eqnarray}}where use has been made of the orthogonality relations of Legendre
polynomials. The integrations now give us the
following expression for $M_{eh}$:
\begin{equation}\label{m00parabolic}
M_{eh}=\frac{1}{2}\frac{e^2}{\pi^{3/2}\epsilon_0\epsilon_r}
\frac{1}{\sqrt{d_e^2+d_h^2}}.
\end{equation}

In a similar manner, we may also approximate the behaviour of $M_{eh}$ in the presence of an external electric field.
For a constant field applied to the dot, the potential in the field direction
(for simplicity say $z$, although the spherical symmetry we assume means all three directions are equivalent) becomes
\begin{equation}\label{electricparabolic}
    V(z)\mapsto V(z)+qFz,
\end{equation}
where $q=-e$ for conduction band electrons, $q=+e$ for holes, and $F$ is the electric field strength. Substituting this into the
Schr\"{o}dinger equation for the $z$-component leads us to a new Schr\"{o}dinger equation that has the
same parabolic potential form:
\begin{equation}\label{newparabolisch}
    \left[-\frac{\hbar^2}{2m^*}\frac{\partial^2}{\partial z'^2}+\frac{1}{2}c_zz'^2\right]\phi(z')
    = E'\phi(z')
\end{equation}
with \begin{equation}\label{electricparabolicx}
\begin{array}{ll}
    z_{e}'=z-eF/c_{e,z} & \hbox{for electrons,} \\
    \\
    z_{h}'=z+eF/c_{h,z} & \hbox{for holes,} \\
    \\
    E_{i}'=E+(eF)^2/2c_{i,z} & \hbox{$i=e, h$.}
\end{array}
\end{equation}
Therefore, electrons and holes are displaced in opposite directions and their
envelope functions are the same as Eq.~\ref{envelopeparabolic}
with $z$ replaced by $z'$. The simplicity of the change in
envelope function with applied electric field is a great advantage
of the parabolic well model, although it should be pointed out
that this same simplicity implies that the charges can continue
separating indefinitely with applied field strength and is
therefore unrealistic at very high fields.

Again, in spherical polar coordinates, the envelope functions in the presence of a field may be written as
\begin{equation}\label{elecenvelopefieldpolars}
  \phi_e({\bf r})=\left(\frac{1}{d_e\sqrt{\pi}}\right)^{3/2}\exp\left({-\frac{({\bf r}-{\bf \hat{k}}eF/c_e)^2}{2d_e^2}}\right)
\end{equation}
for electrons, and
\begin{equation}\label{holeenvelopefieldpolars}
    \phi_h({\bf r})=\left(\frac{1}{d_h\sqrt{\pi}}\right)^{3/2}\exp\left({-\frac{({\bf r}+{\bf \hat{k}}eF/c_h)^2}{2d_h^2}}\right)
\end{equation}
for holes, where $\bf{\hat{k}}$ is the unit vector in the $z$-direction. This time, substituting into Eq.~\ref{directeh} leads to
{\setlength\arraycolsep{0.0em}
\begin{eqnarray}\label{d1field}
M_{eh}&{}={}&C\int\int
\exp\left({-\frac{r_1^2}{d_e^2}}\right)\exp\left({-\frac{r_2^2}{d_h^2}}\right)\nonumber\\
&&{\times}\:\exp\left({{r_1}\alpha\cos{\theta}}\right)\frac{1}{|\bf
r_1-\bf r_2|}\textrm{d} \bf r_1 \textrm{d} \bf r_2,
\end{eqnarray}}
where
{\setlength\arraycolsep{0.0em}
\begin{eqnarray}\label{ehbindingconst}
C&{}={}&\frac{e^2}{4\pi\epsilon_0\epsilon_r}\left(\frac{1}{d_e\sqrt{\pi}}\right)^3\left(\frac{1}{d_h\sqrt{\pi}}\right)^3
\nonumber\\
&&{\times}\:\exp\left[{-\frac{(eF)^2}{d_e^2}\left(\frac{1}{c_e}+\frac{1}{c_h}\right)^2}\right],
\end{eqnarray}}
and
\begin{equation}\label{alpha}
\alpha=\frac{2eF}{d_e^2}\left(\frac{1}{c_e}+\frac{1}{c_h}\right).
\end{equation}
We proceed as before, again making use of Legendre polynomials and their orthogonality relations, and integrate over ${\bf r_2}$ to get
{\setlength\arraycolsep{0.0em}
\begin{eqnarray}\label{errorfunction}
M_{eh}&{}={}&\frac{2\pi^{5/2}Cd_h^3}{\alpha}\int
\exp\left({-\frac{r_1^2}{d_e^2}}\right)\textrm{erf}\left(\frac{r_1}{d_h}\right)\nonumber\\
&&{\times}\:\left[\exp\left({{r_1}\alpha}\right)-\exp\left({-{r_1}\alpha}\right)\right]\textrm{d} r_1.
\end{eqnarray}}For $\alpha\ll1/d_e$ (valid up to fields of order $10^7$ V/m for the small dots considered here), we expand the exponentials in $\alpha$ up to the term in $\alpha^3$ and integrate over $r_1$. Keeping only the terms up to $(d_e\alpha)^2$ order in the resultant expressions gives us an estimate for the suppression of the electron-hole binding energy as an external field is applied:
{\setlength\arraycolsep{0.0em}
\begin{eqnarray}\label{m00field}
M_{eh}&{}=&{}\frac{e^2}{2\pi^{3/2}\epsilon_0\epsilon_r\sqrt{d_e^2+d_h^2}}\nonumber\\
&&{\times}\:\left[1-\frac{e^2F^2}{3\left(d_e^2+d_h^2\right)}\left(\frac{1}{c_e}+\frac{1}{c_h}\right)^2
\right],
\end{eqnarray}}
which reduces to Eq.~\ref{m00parabolic} at $F=0$.

Equations~\ref{electricparabolicx}~and~\ref{m00field} imply a quadratic dependence of the Stark shift (change in exciton energy) on the applied electric field. This has been observed experimentally in a range of QD systems including InGaAs/GaAs,~\cite{rinaldi01} GaAs/GaAlAs,~\cite{heller98} and CdSe/ZnSe.~\cite{seufert01} Furthermore, a theoretical study of an eight-band strain dependent ${\bf k\cdot p}$ Hamiltonian has shown that the quadratic dependence of the ground-state energy on applied field is a good approximation for largely truncated self-assembled quantum dots, although the approximation becomes worse as the dot size increases in the growth direction.~\cite{sheng03}

\section{\label{threea}A Signature of F\"orster Coupled Quantum Dots}
\subsection{\label{theham}Hamiltonian}
We have now characterized the single particle electron and hole
states within a simple QD model, as well as accounting
for the binding energy due to electron-hole coupling within
a dot when estimating the ground state exciton energy.
In this section we shall consider excitons in two coupled QD's and
the Coulomb interactions between them. More specifically, we shall
derive an analytical expression for the strength of the inter-dot
F\"{o}rster coupling. We shall show that this coupling is, under certain conditions, of
dipole-dipole type~\cite{forster59,dexter53} and that it is responsible for
resonant exciton exchange between adjacent QD's. This is a
transfer of energy only, not a tunnelling effect. We are concerned in this paper with bringing excitons within adjacent QD's into resonance. As the Appendix shows, single particle tunnelling is only significant when the energies of the states before and after the tunnelling event are separated by less than the tunnelling energy. This is a different resonant condition to the one considered here and is not fulfilled by the dots over the parameter ranges explored.

Following Ref.~\onlinecite{lovett02a} we write the Hamiltonian
of two interacting QD's in the computational basis
$\{|00\rangle, |01\rangle,|10\rangle,|11\rangle\}$ as ($\hbar =1$)
\begin{equation}\label{eq:Hdot}
\widehat{H}= \left(
\begin{array}{cccc}
\w_0 & 0 & 0 &0 \\
0 & \w_0+\w_2 & V_{\rm F} & 0 \\
0 & V_{\rm F} &  \w_0+\w_1 &0 \\
0 & 0 & 0 &  \w_0+\w_1+\w_2+V_{\rm XX}
\end{array}\right)
\end{equation}
where the off-diagonal F\"orster interaction
is given by $V_{\rm F}$, and the direct Coulomb binding energy
between the two excitons, one on each dot, is on the diagonal and
given by $V_{\rm XX}$.~\cite{biolatti02} The ground state energy is denoted by
$\w_0$, and $\Delta\omega \equiv \w_1-\w_2$ is the difference
between the excitation energy for dot I and that for dot II. These excitation energies and inter-dot interactions are all functions of the applied field $F$. The energies and eigenstates of
this four-level system are given by
\begin{equation}\label{eigen}
    \begin{array}{ll}
      E_{00}= \w_0,& |\Psi_{00}\rangle=|00\rangle \\
      E_{-}=\w_0+\w_1-\frac{\Delta\omega}{2}(1+A), & |\Psi_{-}\rangle=a_1|10\rangle-a_2|01\rangle \\
      E_{+}=\w_0+\w_1-\frac{\Delta\omega}{2}(1-A),  & |\Psi_{+}\rangle=a_1|01\rangle+a_2|10\rangle \\
      E_{11}=\w_0+\w_1+\w_2+V_{\rm XX}, & |\Psi_{11}\rangle=|11\rangle, \\
    \end{array}
\end{equation}
where $A=\sqrt{1+4(V_{\rm F}/\Delta\omega)^2}$, $a_1=\sqrt{(A-1)/2A}$, and $a_2={\rm sgn}(V_{\rm F}\Delta\omega)\sqrt{(A+1)/2A}$ for $|\Delta\omega|>0$.
We can see that $V_{\rm F}$ may cause a mixing of the states $|01\rangle$ and $|10\rangle$
with the result that $|\Psi_{-}\rangle$ and $|\Psi_{+}\rangle$ can
now be entangled states. It is
also straightforward to see that an off-diagonal F\"{o}rster
coupling does indeed correspond to a resonant transfer of energy;
if we begin in the state $|10\rangle$ (exciton on dot I, no exciton
on dot II) this will naturally evolve to a state $|01\rangle$ (no
exciton on dot I, exciton on dot II), in a time given by
$\pi/(2V_{\rm F})$, through the maximally entangled state
$2^{-1/2}\left(|10\rangle+i|01\rangle\right)$. An analogous
behaviour is expected for the initial state $|01\rangle$.

\begin{figure}[t]
\centering
\includegraphics[width=3.2in,height=1.6in]{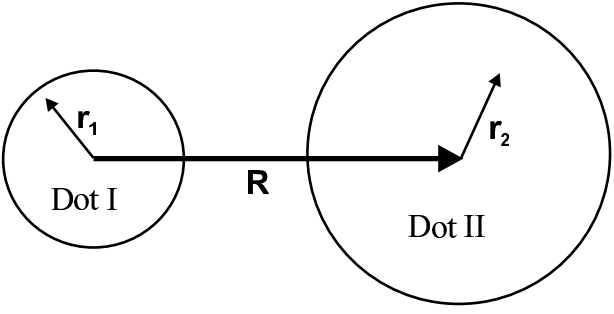}
\caption{Schematic diagram of the interacting two-dot system.}
\label{dots}
\end{figure}

\subsection{Analytical model of the F\"{o}rster interaction}

We shall now calculate the magnitude of the off-diagonal matrix
elements in Eq.~\ref{eq:Hdot} for the parabolic potential model.
We shall see that the behaviour of the
F\"{o}rster interaction due to changes in dot size, composition,
separation, and applied electric fields may be predicted by such
an analytical model. We begin by calculating the form of the matrix
element in the dipole-dipole approximation.

The matrix element we require is that of the Coulomb operator between two single exciton wavefunctions, one located on
each of the two dots. We take our initial
state as representing a conduction band state in dot I, and a valence band state in dot II:
\begin{equation}\label{initialantisymexciton}
    \Psi_{\rm I}= A\left[\psi_n'({\bf r_1,\sigma_1}),\psi_m({\bf r_2,\sigma_2})\right].
\end{equation}
For our final state, we must have a valence band state in dot I and a conduction band state in dot II, given by:
\begin{equation}\label{finalantisymexciton}
    \Psi_{\rm F}= A\left[\psi_n({\bf r_1,\sigma_1}),\psi_m'({\bf
    r_2,\sigma_2})\right].
\end{equation}
The positions ${\bf r_1}$ and ${\bf r_2}$ are now defined from the centres of dot I and dot II respectively (see Fig.~\ref{dots}), and not from the same point as in Eq.~\ref{antisymmetricexciton}, where only a single dot was considered. Note that the time ordering of $\Psi_{\rm I}$ and $\Psi_{\rm F}$ is irrelevant since, as described in Sec.~\ref{threea}, the resonant energy transfer process is reversible.

Therefore, the direct Coulomb matrix element between these two states gives
us
{\setlength\arraycolsep{0.0em}
\begin{eqnarray}\label{forsterintegral}
V_{\rm F}&{}={}&-\frac{e^2}{4\pi\epsilon_0\epsilon_r}\int\int\psi_n'^*({\bf
r_1})\psi_n({\bf r_1})\frac{1}{|\bf R+\bf r_1-\bf r_2|}\nonumber\\
&&{\times}\:\psi_m^*({\bf r_2})\psi_m'({\bf r_2})\textrm{d} \bf r_1 \textrm{d}
\bf r_2 ,
\end{eqnarray}}where we have explicitly included the inter-dot separation $\bf R$.
For $\bf |R|\gg \bf |r_1-\bf r_2|$, which is valid as long as the characteristic sizes of the wavefunctions, $d_{j}$, are small in comparison to $\bf |R|$, we
can follow the procedure of Dexter~\cite{dexter53} and
expand the Coulomb operator in powers of $(\bf r_{1/2}/\bf R)$ up
to second order. Taking the matrix element between $\Psi_{\rm I}$ and
$\Psi_{\rm F}$ leads to
\begin{equation}\label{forsterelement}
    V_{\rm F} = -\frac{e^2}{4\pi\epsilon_0\epsilon_r R^3} \left[\langle {\bf r}_{\rm{I}} \rangle \cdot \langle {\bf r}_{\rm{II}} \rangle
-\frac{3}{R^2} (\langle {\bf r}_{\rm I} \rangle \cdot {\bf
R})(\langle {\bf r}_{\rm II} \rangle \cdot {\bf R})\right].
\end{equation}
We assume the dots are sufficiently separated for there to be no
overlap of envelope functions between dot I and dot II. Therefore,
we do not consider the exchange term of this Coulomb interaction.
The integrals
\begin{equation}\label{positionelement}
\begin{array}{c}
    \langle {\bf r}_{\rm {I}}\rangle=\int\psi_n'^*({\bf r_1}){\bf r_1}\psi_n({\bf r_1})\textrm{d} \bf
    r_1,\\
    \\
    \langle {\bf r}_{\rm {II}}\rangle=\int\psi_m^*({\bf r_2}){\bf r_2}\psi_m'({\bf r_2})\textrm{d} \bf
    r_2,\\
\end{array}
\end{equation}
are taken between an electron and hole ground state centered on dot I and
dot II respectively. Remembering that our wavefunctions are a
product of an envelope function $\phi({\bf r})$ and a Bloch
function $U({\bf r})$ we can make use of their different
periodicities to write~\cite{lovett02}
{\setlength\arraycolsep{0.0em}
\begin{eqnarray}\label{forsterenvelope}
V_{\rm F}&{}={}&-\frac{1}{4\pi\epsilon_0 \epsilon_r R^3} O_{\rm I} O_{\rm II}
\bigg[{\bf d_{\rm{cv(I)}}}\cdot{\bf d_{\rm{cv(II)}}}\nonumber\\
&&{-}\:\frac{3}{R^2} ({\bf d_{\rm{cv(I)}}} \cdot
{\bf R})({\bf d_{\rm{cv(II)}}} \cdot
{\bf R})\bigg].
\end{eqnarray}}
The overlap integrals are defined as
\begin{equation}\label{overlapintegral}
    O=\int_{space}\phi_e({\bf r})\phi_h({\bf r})\textrm{d} \bf
    r,
\end{equation}
with $O_{\rm{I/II}}$ referring to the overlap of the envelope functions for dot I or
dot II respectively (each having a maximum value of unity), and the inter-band dipole matrix elements are defined as
\begin{equation}\label{interbanddipole}
  {\bf d_{\rm{cv}}}=e\int_{cell}U_e({\bf r}){\bf r}U_h({\bf r})\textrm{d} \bf r,
\end{equation}
with ${\bf d_{\rm{cv(I/II)}}}$ referring to dot I or dot II respectively.

We shall not calculate the values of $\bf d_{\rm{cv}}$ here as they are commonly measured experimental quantities (see also
Ref.~\onlinecite{lovett02} for a simple model) and, once the dot
materials have been chosen, are constant contributions to the
F\"{o}rster interaction strength. However, the calculation of $O_{\rm{I/II}}$ is vital in
determining the effects of dot size, shape, and applied electric
fields on the strength of the inter-dot interaction.

\begin{figure}[t]
\centering
\includegraphics[width=3.2in,height=2.8in]{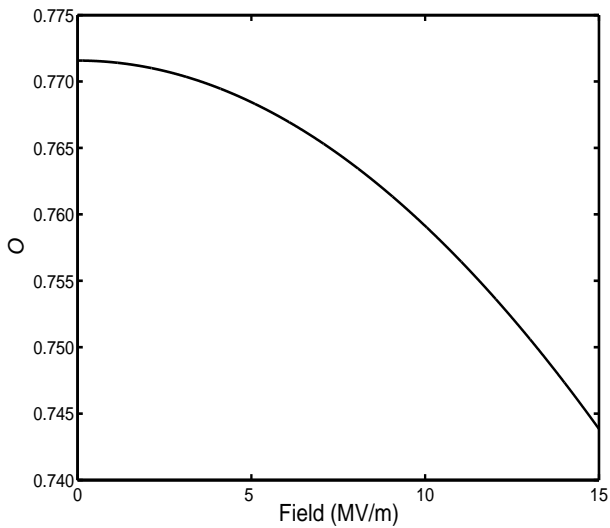}
\caption{
Dependence of the overlap integral of a single parabolic QD as a function of the
in-plane ($x$- direction) electric field strength. Suppression of
the F\"orster interaction results from the reduction in
electron-hole overlap for increasing fields. A 162.5 meV potential at $x=3$ nm from the dot centre for both electrons and holes is used, giving $c=0.00579$ J/$\rm {m}^2$ (see Fig.~\ref{dotparameters}).}
\label{varyf}
\end{figure}

We take the parabolic solutions in Cartesian coordinates from Eq.~\ref{envelopeparabolic}
and also include the effect of a
lateral electric field, which is important in determining how to
suppress the interaction when required, or bring two non-identical dots into resonance. As before,
we shall assume that the
electric field affects only the envelope function part of
the wavefunction; this is valid in the regime where the electric
field never becomes so large that the envelope function varies on
the unit cell scale. This is equivalent to saying that the envelope functions can be decomposed into a
superposition of crystal momentum (${\bf k}$) eigenstates near the band edges, where the Bloch functions are approximately independent of ${\bf k}$.

For a constant field in the lateral direction (say, $x$) the overlap integrals $O_{\rm{I/II}}$ are straightforward to
calculate. Again, for clarity, we take wells identical in all three
directions for both electrons and holes ($d_x=d_y=d_z=d$), leading to
{\setlength\arraycolsep{0.0em}
\begin{eqnarray}\label{elecenvelopeparabolicfield}
  \phi_e({\bf r})&{}={}&\left(\frac{1}{d_e\sqrt{\pi}}\right)^{3/2}\exp\left({-\frac{(x-eF/c_e)^2}{2d_e^2}}\right)
\nonumber\\
&&{\times}\:\exp\left({-\frac{(y^2+z^2)}{2d_e^2}}\right),
\end{eqnarray}}for electrons, and
{\setlength\arraycolsep{0.0em}
\begin{eqnarray}\label{holeenvelopeparabolicfield}
    \phi_h({\bf r})&{}={}&\left(\frac{1}{d_h\sqrt{\pi}}\right)^{3/2}\exp\left({-\frac{(x+eF/c_h)^2}{2d_h^2}}\right)
\nonumber\\
&&{\times}\:\exp\left({-\frac{(y^2+z^2)}{2d_h^2}}\right),
\end{eqnarray}}for holes.
Substituting into Eq.~\ref{overlapintegral} and integrating
results in
\begin{equation}\label{parabolicoverlapresult}
    O=\left(\frac{2d_ed_h}{d_e^2+d_h^2}\right)^{3/2}\exp{\left\{-\frac{e^2F^2(c_e+c_h)^2}{2c_e^2c_h^2(d_e^2+d_h^2)}\right\}}.
\end{equation}
Therefore, in zero applied field the overlap depends only on the
ratio $\left(2d_ed_h/(d_e^2+d_h^2)\right)^{3/2}$. It is worth
noting that if we had chosen an infinite square well potential in
the growth ($z$) direction and parabolic wells in the $x$- and
$y$-directions (as is common in the literature~\cite{krummheuer02}) then
the zero field value would be $\left(2d_ed_h/(d_e^2+d_h^2)\right)$
with the field dependence being exactly the same as in Eq.~\ref{parabolicoverlapresult}.

\begin{figure}[t]
\centering
\includegraphics[width=3.2in,height=2.8in]{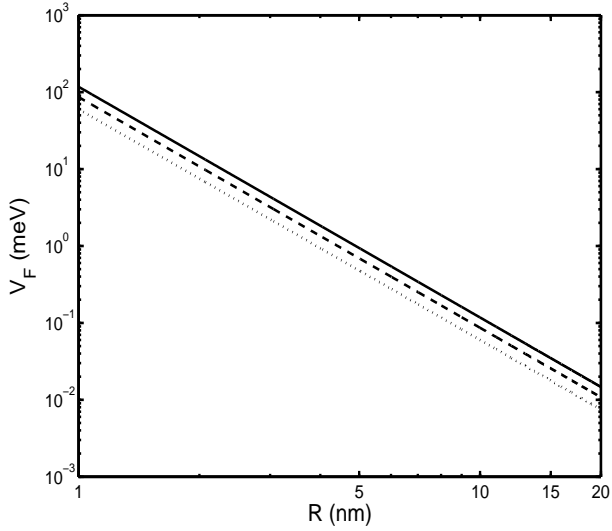}
\caption{
F\"orster interaction strength as a function of dot separation for two identical dots, with $\epsilon_r=12$. Three different values of $e\langle {\bf r}\rangle$ are shown: $7$ e{\AA} (solid line), $6$ e{\AA} (dashed line), and $5$ e{\AA} (dotted line)}
\label{forster}
\end{figure}

\begin{figure*}[t]
\centering
\includegraphics[width=4.5in,height=4.5in]{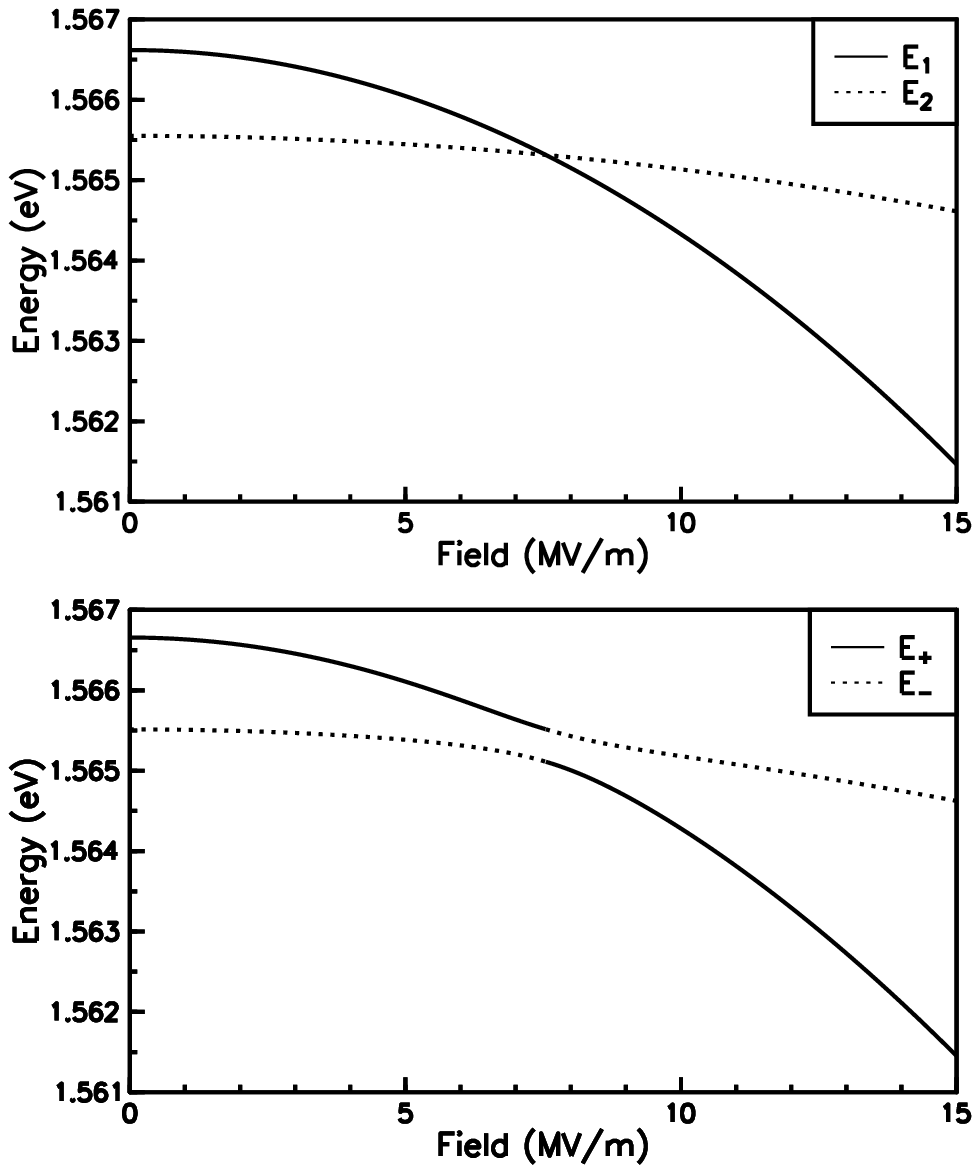}
\caption{(a) Single dot energies as a function of applied electric field. Parameters: $m_e^*=0.04m_0$, $m_h^*=0.45m_0$, $\epsilon_r=12$, and $c_{\rm I}=0.00579$ J/$\rm {m}^2$ (corresponding to a potential of 162.5 meV at a distance of 3~nm from the dot centre) for dot I, for both electrons and holes, and $c_{\rm II}=0.03414$ J/$\rm {m}^2$ (corresponding to a potential of 515.7 meV at a distance of 2.2 nm from the dot centre) for dot II, for both electrons and holes (see Fig.~\ref{dotparameters}). $E^{gap}$ is taken as 1.1 eV for dot I and 0.394 eV for dot II. (b) Energies $E_{-}$ and $E_{+}$ of the coupled dot system demonstrating anticrossing at a field of approximately $7.5\times10^{6}$ V/m. Here, $V_{\rm F}$ has a magnitude of 0.20 meV at zero field, with ${\bf d_{\rm{cv}}}=7$ e{\AA} and $R=7$ nm.}
\label{anticross}
\end{figure*}

Figure~\ref{varyf} shows the suppression of the overlap integrals by an
in-plane electric field ($x$-direction) and therefore the suppression of the F\"orster interaction itself. However, as we shall see in the next section, this does not rule out its observation in coupled dot systems that are tuned to resonance with an external applied field. Indeed, if the inter-dot interaction matrix element is relatively large in zero applied field, then interesting anticrossing behaviour should be observed in the energy spectrum as the system is tuned through resonance. We also note that a suppressed F\"orster coupling can be of benefit to the exciton-exciton dipole interaction quantum computation schemes~\cite{biolatti02} as it ensures an almost purely diagonal interaction between adjacent QD's.

Taking measured values for the transition dipole moment $e\langle {\bf r}\rangle$ allows us to estimate the magnitude of $V_{\rm F}$ between two stacked dots. In CdSe QD's a value for $e\langle {\bf r}\rangle$ of up to $5.2$ e{\AA} has been reported~\cite{crooker02} while for both InGaAs/GaAs and InAs/InGaAs QD's values of approximately $5$-$7$ e{\AA} have been measured.~\cite{silverman03,eliseev00} Considering Eq.~\ref{forsterelement} with a value of $6$ e{\AA} for $e\langle {\bf r}\rangle$, $\epsilon_r=12$ (for InGaAs/GaAs), and inter-dot spacing $R=5$~nm, we obtain an estimate of $0.69$ meV for the F\"orster coupling energy $V_{\rm F}$, certainly large enough to be observed experimentally. This corresponds to a resonant energy transfer time of picosecond order and is therefore interesting as a coupling mechanism for performing quantum logic gates, as it is well within the nanosecond dephasing times~\cite{borri01,birkedal01,bayer02} expected for excitons within QD's (see Sec.~\ref{furtherwork} for further discussion). In Fig.~\ref{forster} the $1/R^3$ dependence of the F\"orster interaction strength is shown for various values of $e\langle {\bf r}\rangle$.

In the next section we shall discuss a signature of the F\"orster interaction that would be observable through photoluminescence measurements.

\subsection{\label{sectionanticrossing}Anticrossings: A signature of F\"orster coupling}

If we consider again Eq.~\ref{eigen} we can see that $|\Psi_{-}\rangle$ and $|\Psi_{+}\rangle$ have a range of forms depending upon the values of $a_1$ and $a_2$. For example, if $\Delta\w\gg V_{\rm F}$, then $A\simeq 1$ which leads to $|a_1|\simeq 0$ and $|a_2|\simeq 1$. Therefore the states $|\Psi_{-}\rangle$ and $|\Psi_{+}\rangle$ are given by $|01\rangle$ and $|10\rangle$ respectively and there is no mixing of the computational basis states. The only way to couple two dots in this case is via the diagonal interaction $V_{\rm XX}$. However, for two dots coupled by $V_{\rm F}$ at resonance ($\Delta\w=0$) we can see from Eq.~\ref{eigen}, and by using $A=\sqrt{(\Delta\w^2+4V_{\rm F}^2)/\Delta\w^2}$, that $|a_1|=|a_2|=1/\sqrt{2}$, $E_{\rm lower}=\w_0+\w_1-|V_{\rm F}|$, and $E_{\rm higher}=\w_0+\w_1+|V_{\rm F}|$.
Furthermore, the two eigenstates $2^{-1/2}(|01\rangle+|10\rangle)$ and $2^{-1/2}(|10\rangle-|01\rangle)$ are both maximally entangled and separated in energy by $2V_{\rm F}$.

Interestingly, we should be able to move between these two cases by bringing two initially non-resonant coupled dots into resonance, for example by the application of a static external electric field. By taking the dots through the resonance, an anticrossing of the energy levels should be observable through photoluminescence measurements. However, the transition from the antisymmetric state to the ground state is not dipole allowed on resonance and should also display a characteristic loss of intensity close to the resonant condition (see Sec.~\ref{decay}).

From Eq.~\ref{electricparabolicx} we see that an external field reduces the energy $E$ for both electrons and holes, with the shift being greater for bigger dots. We therefore consider two coupled dots of different material concentrations, with one of slightly greater dimensions and having a larger band-gap, and in Fig.~\ref{anticross}(a) show that the single dot energy levels cross (for our choice of parameters) as an electric field is applied.  Such a situation is plausible for systems such as InGaAs where dot layers of varying indium content, and hence varying band-gap, may be grown~\cite{zhang01} and it applies directly to the parameters chosen here (see also Sec.~\ref{furtherwork} below). Diagonalising the Hamiltonian (Eq.~\ref{eq:Hdot}) as the field strength $F$ varies gives us a model prediction for the behaviour of the energy levels $E_{-}$ and $E_{+}$ as shown in Fig.~\ref{anticross}(b), where an anticrossing is observed at a field of approximately $7.5\times10^{6}$~V/m. Since $\w_1$, $\w_2$, and $V_{\rm F}$ are all functions of $F$ (as is also shown in Eqs.~\ref{electricparabolicx},~\ref{m00field} and~\ref{parabolicoverlapresult}) there is large scope for finding parameter regimes with interesting behaviour.

An analytical expression for the field strength at resonance can be calculated from the condition $\Delta\w=0$. The total energy of each dot is given by
\begin{equation}\label{omegadot}
E=E^{gap}+\frac{3\hbar}{2}\left(\sqrt{\frac{c}{m_e^*}}+\sqrt{\frac{c}{m_h^*}}\right)-\frac{(eF)^2}{c}-M_{eh},
\end{equation}
for identical potential wells in all directions for both electrons and holes ($c_x=c_y=c_z=c)$. When dot I and dot II are resonant
{\setlength\arraycolsep{0.0em}
\begin{eqnarray}\label{resonantE}
    E_{\rm I}-E_{\rm II}&{}={}& 0 \nonumber\\
&{}={}&\:(E_{\rm I}^{gap}-E_{\rm II}^{gap})+\frac{3\hbar}{2}\Bigg[\sqrt{c_{\rm I}}\left(\frac{1}{\sqrt{m_e^*}}+\frac{1}{\sqrt{m_h^*}}\right)\nonumber\\
&&{-}\:\sqrt{c_{\rm II}}\left(\frac{1}{\sqrt{m_e^*}}+\frac{1}{\sqrt{m_h^*}}\right)\Bigg]\nonumber\\
&&{-}\:(eF)^2\left(\frac{1}{c_{\rm I}}-\frac{1}{c_{\rm II}}\right)-M_{eh_{\rm I}}+M_{eh_{\rm II}},
\end{eqnarray}}
and therefore
{\setlength\arraycolsep{0.0em}
\begin{eqnarray}\label{fsqresonance}
    F^2&{}={}&\frac{1}{\beta e^2}\Bigg\{(E_{\rm I}^{gap}-E_{\rm II}^{gap})\nonumber\\
&&{+}\:\frac{3\hbar}{2}\Bigg[(\sqrt{c_{\rm I}}-\sqrt{c_{\rm II}})\left(\frac{1}{\sqrt{m_e^*}}+\frac{1}{\sqrt{m_h^*}}\right)\Bigg]\nonumber\\
&&{-}\:\frac{e^2}{2\pi^{3/2}\epsilon_0\epsilon_r}\Bigg[\frac{1}{(d_{e_{\rm I}}^2+d_{h_{\rm I}}^2)^{1/2}}
-\frac{1}{(d_{e_{\rm II}}^2+d_{h_{\rm II}}^2)^{1/2}}\Bigg]\Bigg\},\nonumber\\
\end{eqnarray}}
where
{\setlength\arraycolsep{0.0em}
\begin{eqnarray}\label{betaresonance}
\beta &{}={}&\frac{1}{c_{\rm I}}
\left[1-\frac{2e^2}{3\pi^{3/2}\epsilon_0\epsilon_r}
\frac{1}{(d_{e_{\rm I}}^2+d_{h_{\rm I}}^2)^{3/2}}
\frac{1}{c_{\rm I}}\right]\nonumber\\
&&{-}\:\frac{1}{c_{\rm II}}\left[1-\frac{2e^2}{3\pi^{3/2}\epsilon_0\epsilon_r}
\frac{1}{(d_{e_{\rm II}}^2+d_{h_{\rm II}}^2)^{3/2}}
\frac{1}{c_{\rm II}}\right],
\end{eqnarray}}and the states $|\Psi_{-}\rangle$ and $|\Psi_{+}\rangle$ should be maximally entangled at this value of $F$ ($=7.547\times10^{6}$ V/m with the same parameters as for Fig.~\ref{anticross}), with an energy separation equal to $2V_{\rm F}$ (at this field) as stated earlier. Clearly, the experimental observation of an anticrossing as shown in Fig.~\ref{anticross}(b) would be an extremely strong indication of F\"orster coupling between two dots, and also a first indication that entangled states are being produced.

\section{\label{decay}Decay rates and absorption}

We have seen in the previous section that an anticrossing in the energy level structure of two coupled QD's provides a signature of the F\"orster interaction, which should be observable through photoluminescence measurements. However, the scenario considered thus far is idealized in that there is no coupling of the two-dot system to the external environment. Any emission (or absorption) lines associated with coupled-dot transitions will also be broadened due to emission,~\cite{loudon86} and scattering and pure dephasing processes due to exciton-phonon interactions.~\cite{krummheuer02} Experimentally, Bayer and Forchel~\cite{bayer02} have shown that decay processes are dominant at very low temperatures ($\sim2$ K), while other studies have demonstrated good Lorentzian fits to the photoluminescence lineshapes at $5$~K.~\cite{birkedal01,borri03} Hence, we shall limit the discussion here to spontaneous emission (decay) processes.

The spontaneous emission rate for a two-level system (single dot) interacting with a single radiation mode
is usually calculated by considering a quantum mechanical description of the radiation field (see, for example, Ref.~\onlinecite{basu97}), although approaches which consider a classical light field also exist.~\cite{goupalov03} For a dot surrounded by material of approximately the same relative permittivity $\epsilon_r$ the decay rate becomes~\cite{thranhardt02}
\begin{equation}\label{spontfinal}
\Gamma^{sp}=\sqrt{\epsilon_r}\frac{\omega_{10}^{3}|O{\bf d_{\rm{cv}}}|^{2}}{3\pi c^{3}\hbar\epsilon_{0}},
\end{equation}
where $\hbar\omega_{10}$ is the energy difference of the two levels under consideration.
For a typical InGaAs dot we take the parameters $\omega_{10}=1.3$~eV, $O{\bf d_{\rm{cv}}}=6$~e{\AA}, and $\epsilon_r=12$ to give $\Gamma^{sp}=1.04\times10^9$~${\rm s}^{-1}$ or a decay time of $\tau_{decay}=1/\Gamma^{sp}=964$~ps, comparable with experimentally measured exciton lifetimes in this system.~{\cite{birkedal01,borri01,bayer02}}

Here, we are primarily interested in the properties of two interacting dots which form the four-level system considered in Sec.~\ref{theham}. The various decay rates between each level may be calculated in the same manner as for the two-level system previously considered, providing that the changes in transition dipole moments due to the interaction are properly accounted for. We will then be able to predict the typical linewidths that would be observed in experimental measurements of these transitions. We characterize the dipole operator in the computational basis according to which dot the transition occurs within:
\begin{equation}\label{dipolematrix}
\begin{array}{ccccc}
  \langle00| & \langle01| & \langle10| & \langle11| \\
  0 & O_{\rm II}{\bf d_{\rm{cv(II)}}} & O_{\rm I}{\bf d_{\rm{cv(I)}}} & 0 & |00\rangle \\
  O_{\rm II}{\bf d_{\rm{cv(II)}}} & 0 & 0 & O_{\rm I}{\bf d_{\rm{cv(I)}}} & |01\rangle \\
  O_{\rm I}{\bf d_{\rm{cv(I)}}} & 0 & 0 & O_{\rm II}{\bf d_{\rm{cv(II)}}} & |10\rangle \\
  0 & O_{\rm I}{\bf d_{\rm{cv(I)}}} & O_{\rm II}{\bf d_{\rm{cv(II)}}} & 0 & |11\rangle \\
\end{array}.
\end{equation}
Transitions such as $|11\rangle\rightarrow|00\rangle$ have zero dipole moment since the corresponding integral is zero due to the orthogonality of valence and conduction band wavefunctions on each dot:
{\setlength\arraycolsep{0.0em}
\begin{eqnarray}\label{zerodipole}
\langle{\bf r}\rangle&{}={}&\int\psi_n^*({\bf
r_1})\psi_m^*({\bf r_2})({\bf r_1}+{\bf r_2})\psi_n'({\bf
r_1})\psi_m'({\bf r_2})\textrm{d} \bf r_1 \textrm{d}
\bf r_2\nonumber\\
&{}={}&\int\psi_n^*({\bf
r_1}){\bf r_1}\psi_n'({\bf
r_1})\textrm{d} \bf r_1\int\psi_m^*({\bf r_2})\psi_m'({\bf r_2})\textrm{d}
\bf r_2\nonumber\\
&&{+}\:\int\psi_n^*({\bf
r_1})\psi_n'({\bf
r_1})\textrm{d} \bf r_1
\int\psi_m^*({\bf r_2}){\bf r_2}\psi_m'({\bf r_2})\textrm{d}
\bf r_2\nonumber\\
&{}={}&0.
\end{eqnarray}
As a result of Eq.~\ref{dipolematrix}, we may express the dipole moments for general transitions such as $a|01\rangle\pm b|10\rangle\rightarrow|00\rangle$ by
\begin{equation}\label{entangleddmoment}
\langle{\bf r}\rangle_{a,b}=a\langle00|{\bf r}|01\rangle\pm b\langle00|{\bf r}|10\rangle=aO_{\rm II}{\bf d_{\rm{cv(II)}}}\pm bO_{\rm I}{\bf d_{\rm{cv(I)}}},
\end{equation}
which may then be inserted directly into Eq.~\ref{spontfinal}, along with the correct frequencies, to give the corresponding decay rates. In Fig.~\ref{spontemission} we plot $\Gamma^{sp}$ for the two energy curves of Fig.~\ref{anticross}~(b) from Eqs.~\ref{eigen},~\ref{spontfinal} and~\ref{entangleddmoment}, and with the same parameters as Fig.~\ref{anticross}.
\begin{figure}[t]
\centering
\includegraphics[width=3.2in,height=2.8in]{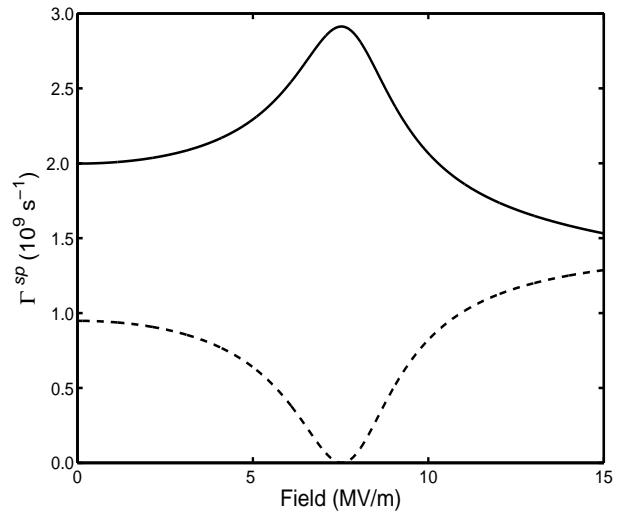}
\caption{Spontaneous emission rates of the coupled-dot energy levels of Fig.~\ref{anticross}. The dashed curve corresponds to the upper curve in Fig.~\ref{anticross}~(b); the solid curve corresponds to the lower curve in Fig.~\ref{anticross}~(b).}
\label{spontemission}
\end{figure}
A special case occurs for identical dots at the anticrossing. Here, the symmetric eigenstate $2^{-1/2}(|01\rangle+|10\rangle)$ has a transition dipole moment to the ground state of $\sqrt{2}O{\bf d_{\rm{cv}}}$ ($O_{\rm I}=O_{\rm II}=O$, ${\bf d_{\rm{cv(I)}}}={\bf d_{\rm{cv(II)}}}={\bf d_{\rm{cv}}}$) and hence a decay rate of twice that expected for a single dot. However, the antisymmetric eigenstate $2^{-1/2}(|01\rangle-|10\rangle)$ has no transition dipole moment and consequently no spontaneous emission rate in the dipole approximation. This is otherwise known as a ``dark" state and any spectral line corresponding to this transition will display a characteristic narrowing and loss of intensity as the anticrossing is approached.

This effect may be studied in more detail by considering the absorption lineshape of each transition (within the rotating-wave approximation):\cite{loudon86,haug94}
\begin{equation}\label{absorptionfinal}
\alpha(\omega)=\frac{\omega e^2|\langle{\bf r}\rangle|^2}{c\eta\hbar\epsilon_0}\frac{(\Gamma^{sp}/2)}{(\omega_{10}-\omega)^2+(\Gamma^{sp}/2)^2},
\end{equation}
where $\eta$ is the refractive index.
Here, the only line broadening which is accounted for is due to spontaneous emission. This leads to absorption lines with a Lorentzian dependence on frequency, and a full width at half maximum given by $\Gamma^{sp}$. Although other mechanisms may also broaden the lines, for example ``pure" dephasing due to exciton-phonon interactions as mentioned earlier, these processes can usually be reduced, in our case by cooling the system.~{\cite{borri03}} However, it is difficult to reduce the spontaneous emission rate of a given transition. Therefore, the linewidth $\Gamma^{sp}$ is the minimum achievable from any standard dot sample and hence it is vitally important to ensure that any effects we wish to observe will not be masked by its presence.

\begin{figure}[t]
\centering
\includegraphics[width=3.3in,height=6.2in]{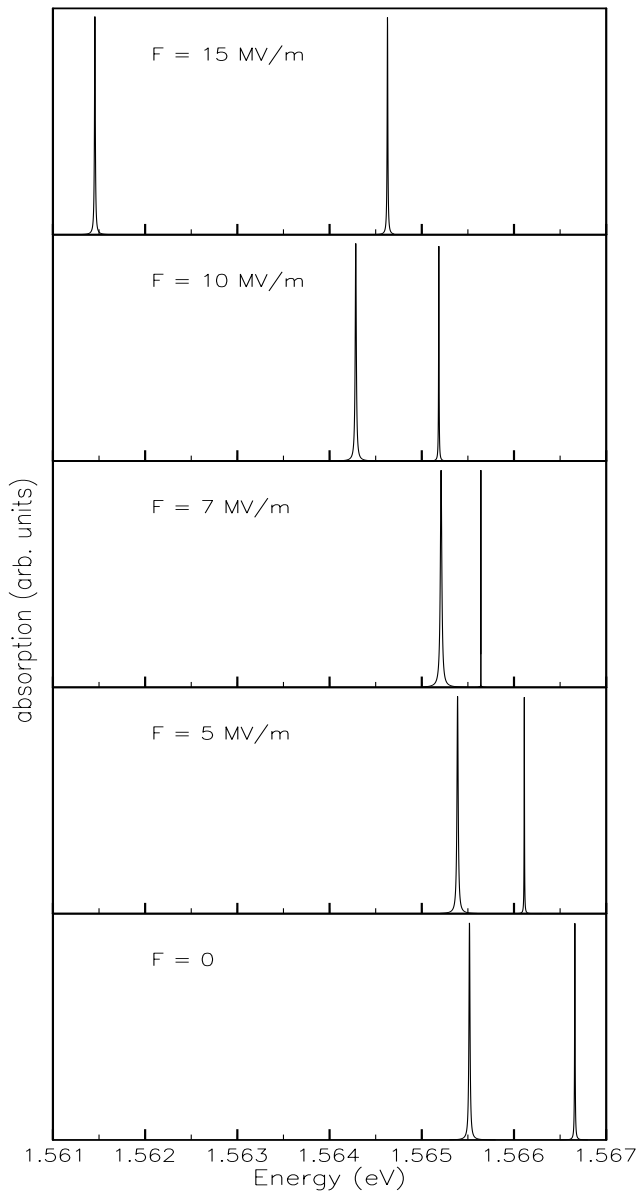}
\caption{Series of simulated absorption spectra of the energy levels in Fig.~\ref{anticross}(b) at fields of $F=0,5,7,10,15$~MV/m. These lines have been artificially broadened by a factor of 10.}
\label{absorption}
\end{figure}

Plotted in Fig.~\ref{absorption} are the absorption spectra of the two energy levels in Fig.~\ref{anticross}(b) ($E_+$ and $E_-$) at fields of $0,5,7,10,15$~MV/m respectively, calculated from Eq.~\ref{absorptionfinal} with the same parameters as for Fig.~\ref{anticross} and $\eta=3.46$. The spontaneous emission rates are calculated from Eq.~\ref{spontfinal} and the peaks have been artificially broadened by a factor of 10 to exaggerate their characteristic features in the changing applied field. As the field is increased the two peaks shift to lower energies with the initially higher energy line (corresponding to $E_+$) shifting by a greater amount so that their separation reduces. The width of the lower energy line ($E_-$) increases as the anticrossing point ($F=7.5$~MV/m) is approached signifying its increasing decay rate (see Eq.~\ref{entangleddmoment}). On the other hand, the width of the $E_+$ line decreases as it approaches a dark state, and the area underneath the curve is reduced. This would correspond to a lowering of intensity of this line in a photoluminescence experiment. We can see that the separation of the lines due to the F\"orster interaction
($2V_{F}\sim0.4$~meV) is well resolved in the presence of radiative broadening at low temperatures, and should be so even if the lines have extra broadening due to exciton-phonon interactions. In fact, sharp emission lines of approximately $0.1$~meV width have been obtained from single InGaAs QD's at a temperature of $100$~K,~\cite{bayer02} indicating that up to this temperature at least, the anticrossing effect should still be observable. At very high fields ($F=10-15$~MV/m) beyond the anticrossing point the two lines once again become well separated and eventually have similar widths, indicating that the states $|01\rangle$ and $|10\rangle$ are now only weakly coupled.

\section{\label{furtherwork}Connection to Quantum Information Processing}

The experimental observation through photoluminescence of an anticrossing of the type above would be a significant step towards proof-of-principle experiments; although it could be a difficult experiment to perform, this method may well yield results more quickly than an attempt at coherent control on dot systems coupled in this way.

The main question here is the feasibility of bringing two dots into resonance using a static external electric field. As has been mentioned above, and can be seen in Fig.~\ref{anticross}(a), a larger dot experiences a larger shift in its energy levels due to the applied field than a smaller one. However, all other parameters being equal, a larger dot also has slightly lower energy at zero field (which is not the case in Fig.~\ref{anticross}). Therefore, a way is needed of increasing the initial energy of the larger dot relative to the smaller one. This could be realised by using layers of different materials (or material concentrations) to alter the band-gap within each dot; other methods such as exploiting different dot geometries or applying a local strain or electric field gradient should also be explored. In Ref.~\onlinecite{rinaldi01} field gradients close to $20$~(MV/m)/$\mu$m were generated, with Stark shifts of approximately $2$~meV obtained in a field of $0.2$~MV/m. Hence, similar dots of $1-2$~meV initial energy separation, and placed $7$~nm apart, could be brought into resonance by this method.
Nitride QD's could also offer a promising approach since their strong piezoelectric fields allow the possibility of an external field shifting their energy levels
towards each other.

We have also shown that it should be possible to engineer
nanostructures such that the off-diagonal F\"{o}rster interaction
between a pair of QD's is of the required strength to make
it interesting for quantum computation.
Once a measurement of this coupling strength is made, the next
logical step is to attempt to controllably entangle the excitonic
states of two interacting dots, leading on to a demonstration of a
simple quantum logic gate such as the controlled-NOT (CNOT).
Although, on resonance, the states $|01\rangle$ and $|10\rangle$
naturally evolve into maximally entangled states after a time
$\pi/(4V_{\rm F})$, their initialization requires the inter-dot
interaction to be suppressed. Furthermore, the generation of a logic gate
such as the CNOT requires single qubit operations on both dots, as
well as periods of interaction.

By switching to a pseudo-spin description of our excitonic qubit
we can immediately consider a previously known operation sequence for the realization of a CNOT gate.
Defining $|\uparrow_z\rangle\equiv|0\rangle$ and
$|\downarrow_z\rangle\equiv|1\rangle$, we can see from Eq.~\ref{eq:Hdot} that the off-diagonal terms can be expressed
as\begin{equation}\label{spinforster}
    H_{\rm F}=\frac{V_{\rm F}}{2}\left(\sigma_{x_1}\sigma_{x_2}+\sigma_{y_1}\sigma_{y_2}\right),
\end{equation}
with \begin{equation}\label{sx}
\sigma_x =\left(\begin{array}{cc}
  0\phantom{.} & 1 \\
  1\phantom{.} & 0 \\
\end{array}\right)  {\rm{and}}
~\sigma_y =\left(\begin{array}{cc}
  0\phantom{.} & -i \\
  i\phantom{.} & 0 \\
\end{array}\right),
\end{equation}
being two of the Pauli spin matrices. This \textit{XY} type
Hamiltonian has been studied in the literature for various systems~\cite{siewert00,imamoglu99} and if the two interacting qubits are left for a time
$t=\pi/(2V_F)$ solely under its influence then an iSWAP gate
will be executed:~\cite{schuch03}\begin{equation}\label{iswap}
    \textrm{iSWAP}=\left(\begin{array}{cccc}
      1\phantom{..} & 0\phantom{..} & 0\phantom{..} & 0 \\
      0\phantom{..} & 0\phantom{..} & i\phantom{..} & 0 \\
      0\phantom{..} & i\phantom{..} & 0\phantom{..} & 0 \\
      0\phantom{..} & 0\phantom{..} & 0\phantom{..} & 1 \\
    \end{array}\right),
\end{equation} in the basis $\{|00\rangle,
|01\rangle,|10\rangle,|11\rangle\}$.
Two iSWAP operations may be concatenated with single qubit
operations to form the more familiar CNOT gate:
{\setlength\arraycolsep{0.0em}
\begin{eqnarray}\label{xycnot}
\textrm{CNOT}&{}={}&\left(\frac{\pi}{2}\right)_{x_2}\left(\frac{\pi}{2}\right)_{z_2}\left(-\frac{\pi}{2}\right)_{z_1}
(\textrm{iSWAP})\left(\frac{\pi}{2}\right)_{x_1}\nonumber\\
&&{\times}\:(\textrm{iSWAP})\left(\frac{\pi}{2}\right)_{z_2},
\end{eqnarray}}where $(\pm\pi/2)_{{l}_{m}}$ are single pseudo-spin rotations
of $\pm\pi/2$ about the $l$ axis of spin $m$, for $l=x,z$ and $m=1,2$.
Schuch and Siewert~\cite{schuch03} have also shown that the CNOT and SWAP
operations may be combined when using an $XY$ interaction to
produce more efficient quantum circuits. Furthermore, the iSWAP operation is an entangling gate and is therefore sufficient for universal quantum computation provided that fast local unitary operations are available. In fact, for systems exhibiting an $XY$ interaction, the iSWAP operation constitutes the natural gate choice when implementing efficient quantum circuits.

To perform a gate such as the CNOT outlined above we must be able
to control the interaction between our two qubits so that we can
effectively switch it off for the duration of the single qubit manipulations.
For the case of excitonic qubits, coupled via the F\"{o}rster
mechanism, the most sensible way to proceed is to consider two
initially non-resonant QD's with negligible energy
transfer. Single qubit operations can then be achieved with
external laser pulses by inducing Rabi oscillations within each
dot.~\cite{kama01} As each dot will have a different excitation
energy, we may address them individually by choosing the
appropriate frequency. Two periods of free evolution under the
interaction Hamiltonian (Eq.~\ref{spinforster}) are also required; applying a suitably selected detuned
pulse to both non-resonant dots will bring them into resonance via
the optical Stark effect.~\cite{cohen92,nazir04} We then allow resonant energy
transfer to occur for a time $t=\pi/(2V_{\rm F})$ producing an iSWAP
operation. The detuned pulse is then stopped and single qubit
manipulations may be induced as before.

Figure~\ref{anticross}(b) provides a nice visualisation of the whole process. We must non-adiabatically switch between the two regimes of zero field, where the dots are effectively uncoupled, and the resonant point where the dots interact. It is our hope that this is achievable through the optical Stark effect, and we speculate that this all optical approach may have the potential to allow gates to be performed well within the limits set by the
nanosecond dephasing times experimentally observed.

To summarise, we have analytically calculated the magnitude of the F\"orster energy transfer between a pair of generic QD's and investigated its effect on their energy level structure. We have proposed a simple experiment which provides a signature of the interaction and an estimate of its strength, and have also discussed its possible application to quantum information processing.

\section*{Acknowledgments}
AN, BWL, and JHR are supported by EPSRC (BWL and JHR as part of the Foresight
LINK Award {\it Nanoelectronics at the Quantum Edge, {\it www.nanotech.org}}). BWL thanks
St Anne's College for support. SDB acknowledges support from the E.U. NANOMAGIQC project
(Contract no. IST-2001-33186); We thank
T.~P.~Spiller, W.~J.~Munro, S.~C.~Benjamin and R.~A.~Taylor for stimulating
discussions.

\appendix*
\section{Tunneling}
\label{appendix:a}

To outline the effect of electron and hole tunneling on the exciton states used in this paper, we consider here the  Hamiltonian for an electron-hole pair in a double-dot. The basis we use is constructed of products of the electron and hole single-particle states $\{|e_{\rm I}h_{\rm I}\rangle,|e_{\rm I}h_{\rm II}\rangle,|e_{\rm II}h_{\rm I}\rangle,|e_{\rm II}h_{\rm II}\rangle\}$, which gives
\begin{equation}\label{sparticleH}
H=\left(%
\begin{array}{cccc}
  E_{e_{\rm I}h_{\rm I}} & t_h & t_e & V_{\rm F} \\
  t_h & E_{e_{\rm I}h_{\rm II}} & 0 & t_e \\
  t_e & 0 & E_{e_{\rm II}h_{\rm I}} & t_h \\
  V_{\rm F} & t_e & t_h & E_{e_{\rm II}h_{\rm II}} \\
 \end{array}%
\right),\end{equation}
where $E_{e_{n}h_{m}}=E_{e_n}+E_{h_m}-M_{e_nh_m}$, with $n,m={\rm I,II}$ for dot I and dot II respectively. $M_{e_nh_m}$ is the direct Coulomb binding energy between the electron and hole on dot $n$ and $m$ respectively, and the band-gap energy has been absorbed into the electron energy $E_{e_n}$ by setting the energy zero to be at the top of the valence band. $V_{\rm F}$ is the F\"orster interaction strength, and $t_{e(h)}$ is the electron (hole) tunneling matrix element.

We consider first the simple case of two identical dots coupled to one another; this means setting $E_{e_{\rm I}}=E_{e_{\rm II}}\equiv E_e$, $E_{h_{\rm I}}=E_{h_{\rm II}}\equiv E_h$, $M_{e_{\rm I}h_{\rm I}}=M_{e_{\rm II}h_{\rm II}}\equiv M_{eh}$, and $M_{e_{\rm I}h_{\rm II}}=M_{e_{\rm II}h_{\rm I}}\equiv M'_{eh}$. Subtracting $E_e+E_h-M_{eh}$ from the diagonal of Eq.~{\ref{sparticleH}} gives
\begin{equation}\label{identicaldotsH}
H=\left(%
\begin{array}{cccc}
  0\phantom{.} & t_h & t_e\phantom{.} & V_{\rm F} \\
  t_h\phantom{.} & M_{eh}-M'_{eh} & 0\phantom{.} & t_e \\
  t_e\phantom{.} & 0 & M_{eh}-M'_{eh}\phantom{.} & t_h \\
  V_{\rm F}\phantom{.} & t_e & t_h\phantom{.} & 0 \\
 \end{array}%
\right).\end{equation}
We would like to isolate the $\{|e_{\rm I}h_{\rm I}\rangle,|e_{\rm II}h_{\rm II}\rangle\}$ subspace, as it is composed of single exciton states on each of the two dots. These states are exactly the ones that are relevant for the computational basis introduced in Sec.~\ref{intro}. Any leakage from this subspace, potentially due to tunnel couplings to the states $|e_{\rm I}h_{\rm II}\rangle$ and $|e_{\rm II}h_{\rm I}\rangle$, could be a source of error for the signature and schemes presented here and in Refs.~\onlinecite{lovett02,lovett02a} and must be minimized.
However, under the condition
\begin{equation}\label{appendcond}
|M_{eh}-M'_{eh}|\gg |t_e|,|t_h|,
\end{equation}
we may use degenerate perturbation theory on Eq.~\ref{identicaldotsH} to give
\begin{equation}\label{iddotsHsub}
H_{eff}=\left(%
\begin{array}{cc}
-\frac{t_e^2+t_h^2}{M_{eh}-M'_{eh}} & V_{\rm F}-\frac{2t_et_h}{M_{eh}-M'_{eh}} \\
 V_{\rm F}-\frac{2t_et_h}{M_{eh}-M'_{eh}} & -\frac{t_e^2+t_h^2}{M_{eh}-M'_{eh}} \\
 \end{array}%
\right),\end{equation}
in the $\{|e_{\rm I}h_{\rm I}\rangle,|e_{\rm II}h_{\rm II}\rangle\}$ subspace.
Hence, the states $|e_{\rm I}h_{\rm I}\rangle$ and $|e_{\rm II}h_{\rm II}\rangle$ are still resonantly coupled in the presence of tunneling as long as Eq.~{\ref{appendcond}} is satisfied. These conditions are better satisfied as the inter-dot separation increases (tunneling elements consequently reduce, as does $M'_{eh}$ so that $|M_{eh}-M'_{eh}|$ becomes larger), and as dot confinement increases (tunneling elements reduce, $M_{eh}$ increases so that $|M_{eh}-M'_{eh}|$ again becomes larger). Furthermore, corrections to the eigenstates $|\chi_{\pm}\rangle\equiv2^{-1/2}(|e_{\rm I}h_{\rm I}\rangle\pm|e_{\rm II}h_{\rm II}\rangle)$ due to mixing with states outside the subspace will be small since they are weighted by factors of $t_{e(h)}/(M'_{eh}-M_{eh})$, to first order, from the perturbation theory.

The regime in which Eq.~{\ref{identicaldotsH}} is valid is not necessarily the ideal one for minimizing the effect of tunneling, while exploiting resonant exciton interactions, as two non-identical dots may also be brought into resonance (see Section~\ref{sectionanticrossing}). In this case, $E_{e_{\rm I}}+E_{h_{\rm I}}-M_{e_{\rm I}h_{\rm I}}=E_{e_{\rm II}}+E_{h_{\rm II}}-M_{e_{\rm II}h_{\rm II}}\equiv E$ on resonance. Subtracting $E$ from the diagonal of Eq.~{\ref{sparticleH}} gives
\begin{equation}\label{nonidenticaldotsH}
H=\left(%
\begin{array}{cccc}
  0\phantom{.} & t_h & t_e\phantom{.} & V_{\rm F} \\
  t_h\phantom{.} & \Delta E_h+ \Delta M_h & 0\phantom{.} & t_e \\
  t_e\phantom{.} & 0 & \Delta E_e+\Delta M_e\phantom{.} & t_h \\
  V_{\rm F}\phantom{.} & t_e & t_h\phantom{.} & 0 \\
 \end{array}%
\right),\end{equation} where $\Delta E_i=E_{i_{\rm II}}-E_{i_{\rm I}}$, for $i=e,h$, and $\Delta M_h=M_{e_{\rm I}h_{\rm I}}-M_{e_{\rm I}h_{\rm II}}$, $\Delta M_e=M_{e_{\rm I}h_{\rm I}}-M_{e_{\rm II}h_{\rm I}}$.
The dots must now satisfy the modified condition
\begin{equation}\label{appendcond2}
min(|\Delta E_h+\Delta M_h|,|\Delta E_e+\Delta M_e|)\gg |t_e|,|t_h|,
\end{equation}
in order for tunneling to be neglected, with the unwanted states $|e_{\rm I}h_{\rm II}\rangle$ and $|e_{\rm II}h_{\rm I}\rangle$ weighted by a factors of magnitude
\begin{equation}\label{weight1}
\frac{t_{e(h)}}{|\Delta E_h+\Delta M_h|},
\end{equation}
and
\begin{equation}\label{weight2}
\frac{t_{e(h)}}{|\Delta E_e+\Delta M_e|},
\end{equation}
to first order in a perturbation expansion. Again, tunneling will be suppressed as dot separation and confinement increases.



\end{document}